\documentclass[journal,onecolumn,12pt,draftclsnofoot]{IEEEtran}
\usepackage{graphicx}
\usepackage[noend]{algpseudocode}
\usepackage{algorithmicx,algorithm}
\usepackage{amsmath}
\usepackage{titlesec}
\usepackage{multirow}  
\usepackage{amsthm,amsmath,amssymb}
\usepackage{graphicx}
\usepackage{float}
\usepackage{subfigure}
\usepackage{mathrsfs}
\usepackage{url}
\usepackage{array}
\newcolumntype{C}[1]{>{\centering}p{#1}}
\usepackage{makecell}
\usepackage{diagbox}
\usepackage{amsmath}
\usepackage{gensymb}
\usepackage{cite}
\usepackage{graphicx}
\usepackage{subfigure}
\usepackage{color}

\ifCLASSOPTIONcompsoc
 \usepackage[caption=false,font=normalsize,labelfont=sf,textfont=sf]{subfig}
\else
 \usepackage[caption=false,font=footnotesize]{subfig}
\fi

\UseRawInputEncoding
\begin{document}
\linespread{1.67}
\vspace{-15mm}
\title{Reconfigurable Intelligent Surface for Near Field Communications: Beamforming and Sensing}

\author{Yuhua Jiang, Feifei Gao, Mengnan Jian, Shun Zhang, and Wei Zhang

\thanks{Y. Jiang, and F. Gao are with Institute for Artificial Intelligence, Tsinghua University (THUAI), 
State Key Lab of Intelligent Technologies and Systems，Tsinghua University, 
Beijing National Research Center for Information Science and Technology (BNRist), Beijing, P.R. China (email:  jiangyh20@mails.tsinghua.edu.cn, feifeigao@ieee.org).
M. Jian is with the State Key Laboratory of Mobile Network and Mobile Multimedia Technology, Shenzhen 518055, China, and also with Algorithm Dept., Wireless Product R$\&$D Institute, ZTE Corporation, Shenzhen 518057, China.
S. Zhang is with the State Key
Laboratory of Integrated Services Networks, Xidian University, Xi’an 710071,
P. R. China (e-mail: zhangshunsdu@xidian.edu.cn).
W. Zhang is with the School of Electrical Engineering and Telecommunications, University of New South Wales, Sydney, NSW 2052, Australia (e-mail: w.zhang@unsw.edu.au).}
}
\maketitle
\vspace{-15mm}
\begin{abstract}
Reconfigurable intelligent surface (RIS) can improve the
communications between a source and a destination. Recently, continuous aperture RIS is proved to have better communication performance than discrete aperture RIS and has received much attention. However, the conventional continuous aperture RIS is designed to convert the incoming planar waves into the outgoing planar waves, which is not the optimal reflecting scheme when the receiver is not a planar array and is located in the near field of the RIS.
In this paper, we consider two types of receivers in the radiating near field of the RIS:  (1) when the receiver is equipped with a uniform linear array (ULA), we design RIS coefficient to convert planar waves into cylindrical waves; (2) when the receiver is equipped with a single antenna, we design RIS coefficient to convert planar waves into spherical waves.
\textcolor{black}{
We then propose the maximum likelihood (ML) method and the focal scanning (FS) method to sense the location  of the receiver based on the  analytic expression of the reflection coefficient, and derive the corresponding position error bound (PEB). 
Simulation results demonstrate that the proposed scheme can reduce energy leakage and thus enlarge the channel capacity compared to the conventional scheme. Moreover, the location of the receiver could be accurately sensed by the ML method with large computation complexity or be roughly sensed by the FS method with small computation complexity.
}

\end{abstract}

\IEEEpeerreviewmaketitle

\section{Introduction}
The enhancement of wireless connectivity in the last
decades has radically changed the way humans perceive
and interact with the surroundings. However, the interfacing network infrastructure has been mostly confined at
rooftops and distant-away serving sites.  
Recently, the idea of improving
wireless networks by means of relays has been renovated
through the concept of low-cost smart mirrors. As evidence,
nowadays’ scientific literature is full of appellatives such as
 large intelligent surfaces
(LIS) and reconfigurable intelligent surfaces (RIS) \cite{9384499,9149203,8746155,9145334}.
\textcolor{black}{The difference between RIS and LIS is that LIS can both generate and reflect electromagnetic waves, while RIS is only used to reflect electromagnetic waves.
In particular, RIS has been
deemed as a promising technology for the future wireless
communication networks, due to its capability of adjusting the channel environment \cite{1}, \cite{reviewer}.}


Practical RIS consists of a large number of discrete programmable sub-wavelength elements, \textit{i.e.}, unit cells or meta atoms \cite{RIS}, \cite{9343768}.
For instance, a properly designed unit cell phase distribution across the surface enables the RIS to alter the direction of the wavefront of the reflected wave, thereby realizing the generalized Snell’s law \cite{3}. 
RIS  makes the wireless
environment controllable and programmable, and thus brings unprecedented new opportunities to enhance the performance
of wireless communication systems \cite{9221372,9495362,9416239}.
The channel along the path from the base
station (BS) to RIS and then to the user is modeled as the
cascading convolution of the BS-to-RIS
channel, the diagonal phase shift matrix of the RIS, and the
RIS-to-user channel \cite{2,via,9464314}.

However, discrete elements of the RIS result in discrete phase shifts on the RIS, which may influence the desired reflection property such as the reflection angle, the anomalous reflection, the path loss, and the interference intensity of the scattered waves \cite{be1}, \cite{8466374}. \textcolor{black}{RISs with the continuous aperture have aroused much research attention recently due to its benefits in realizing more sophisticated anomalous reflection and increasing the achievable rate \cite{9424177,9110889,passive}. }
Existing works on the path loss of the continuous RIS-aided communication systems could be classified
into two categories: the antenna theory-based model
and the electromagnetic-theory based model.
For the antenna theory-based model, each RIS element is
equivalent to a small antenna with specific power radiation
pattern, and the Friis Transmission Equation is used to calculate the path loss \cite{3}. In \cite{3}, the authors prove that
the received power at the target receiver grows quadratically
with the RIS elements’ number in the far field. 
For the electromagnetic theory-based model, a boundary
condition problem is formulated  by considering the RIS as a
holographic surface, on which the reflection coefficient at
each point can be designed, and the scattered fields could be
modeled using electromagnetic theorems \cite{5,7,France}. The main advantage of the electromagnetic theory-based
model is that it is derived
from Maxwell equations and underlines
the electromagnetic nature of the RIS more clearly \cite{9247315}, \cite{9120479}.

However, in most literature on the continuous RIS, the transmit antenna and the receiver antenna are both considered in the far field, and the RIS is designed to convert the incident planar wave into the reflected planar wave with an arbitrary angle \cite{9424177}, \cite{9594786}.
For near-field planar arrays, the design of the RIS in \cite{9424177} and \cite{9594786} is still feasible because planar waves are still capable of concentrating power on the receiver.  
However, when the receiver is equipped with uniform linear array (ULA) or a single antenna and stays in the near field of the RIS, the reflected planar wave can not converge on the receiver, which may cause serious energy leakage and lead to poor channel capacity. 
In order to improve the focusing property, the RIS should reflect cylindrical waves or spherical waves, where the axis of the  cylinder is the focal line or the centre of sphere is the focal point. Nevertheless, the price paid is that  the position of focal line or focal point should be known before the near field RIS design.
Therefore, the location of the receiver needs to be sensed in the first place. To the best of the authors' knowledge, there is no previous work exploring how to focus energy on the ULA or the single antenna and how to sense the location of the receiver in the near field of the continuous aperture RIS.

In this paper, we propose a continuous RIS scheme  to convert planar waves into cylindrical waves or spherical waves such that the energy is concentrated on the ULA or the single antenna within the near field region. We derive the analytical expression of the reflection coefficient of the RIS to radiate cylindrical or spherical waves. 
With cylindrical or spherical wave radiation, the power received by the receiver is a  function highly related to its location.
\textcolor{black}{
 At the same time, we propose the maximum likelihood (ML) method and the focal scanning (FS) method to sense the location  of the receiver based on the  analytic expression of the reflection coefficient, and derive the corresponding position error bound (PEB). 
Simulation results demonstrate that the proposed scheme can reduce energy leakage and thus enlarge the channel capacity compared to the conventional scheme. Moreover, the location of the receiver could be accurately sensed by the ML method with large computation complexity or be roughly sensed by the FS method with small computation complexity.
}


The rest of this paper is organized as follows. Section~\uppercase\expandafter{\romannumeral2}
presents the system model and the problem formulation.
Section~\uppercase\expandafter{\romannumeral3} derives the reflected fields via the induction theorem.
Section \uppercase\expandafter{\romannumeral4} describes the reflection coefficient design criterion.
\textcolor{black}{
Section \uppercase\expandafter{\romannumeral5} presents the ML and FS methods to sense the location  of the receiver and the corresponding PEB.
}
Section \uppercase\expandafter{\romannumeral6} provides the numerical simulation results, and
Section \uppercase\expandafter{\romannumeral7} draws the conclusions.

Notations: Boldface denotes vector or matrix;  $j$ corresponds to the imaginary unit; $(\cdot)^H$ , $(\cdot)^T$, and $(\cdot)^*$ represent the Hermitian, transpose, and conjugate, respectively; $\circ$ represents the Hadamard product operator;
$\left|\mathbf{a}\right|$ denotes the vector composed of lengths of each element in complex vector $\mathbf{a}$; $\left\Vert\mathbf{a}\right\Vert$ denotes 2-norm of the vector $\mathbf{a}$. 

\section{System Model}
We consider the RIS as a rectangular plate with length $a$ in $y$-axis  and length $b$ in $x$-axis, located in the horizontal plane. Suppose the wave number is $k=2\pi/\lambda$ where $\lambda$ is the wavelength. We consider continuous-aperture RIS instead of discrete-aperture RIS for the benefits to achieve more sophisticated anomalous reflection and to increase the transmission rate \cite{9424177}, \cite{9110889}. We assume that the RIS can realize a continuous reflection with coefficient function $\Gamma(x, y) = \tau(x,y) e^{j\beta(x,y)}$, where $\tau(x,y)$ is the amplitude coefficient and $\beta(x, y)$
is the phase shift at the point $(x, y, 0)$ on the RIS.

A point source in the far field is radiating
a linearly polarized electromagnetic wave. We assume that
the curvature of the incident wave front over the dimensions
of the RIS can be neglected. Suppose the incident wave is parallel to the $yz$ plane, and the receiver is in the radiating near field (Fresnel) region. Then the reactive fields are negligible compared with the radiating fields. Let $d$ denote the distance between the center of the antenna and the center of RIS. Then the radiating near field region (RNFR) is given by \cite{soft}:
\begin{align}
0.62 \sqrt{\frac{(a^2+b^2)^{3/2}}{\lambda}}<d<\frac{2 (a^2+b^2)}{\lambda}.
\end{align}
The upper bound of the region $\frac{2 (a^2+b^2)}{\lambda}$ is called the Rayleigh Distance. Note that depending on the values of $a$, $b$ and $\lambda$, the RNFR may or may not exist.\footnote{The propose scheme can be easily extended to the case when source is in the near field while the receiver is in the far field, or when both terminals are in the near field.}

\subsection{Channel Model}
\subsubsection{When The Receiver Is Equipped With ULA}
Suppose the receiver is equipped with ULA of $M$ antennas, and the length of the ULA is $L$. As one of the first works studying the electromagnetic model, we assume there is only the line of sight (LOS) path from the source to the RIS and from the RIS to the destination.\footnote{This is also a typical scenario in case of mmWave transmission.} Since the transmit antenna is in the far field, it could be regarded as a point source. The received signal can be written as \cite{France}, \cite{next}

\begin{equation}
\mathbf{y}=\mathbf{h} s+\mathbf{n},
\end{equation}
where $s$ is the transmitted signal, $\mathbf{n}$  is the additive noise, and $\mathbf{h}$ is the overall channel from the transmitter to RIS and RIS to the receiver. .

According to the Huygens-Fresnel principle, 
every point on the RIS is a new source of the spherical wavelets, and the secondary wavelets radiating from different points mutually interfere. The sum of these spherical wavelets generates the wavefront propagating from the RIS to the receiver.
The energy of the wave emitted by a point source falls off as the inverse square of the distance travelled, and thus the amplitude falls off as the inverse of the distance. The complex amplitude of the electric field intensity at the point $(x,y,0)$ is given by
\begin{align}
E(x,y)&=\frac{c e^{j k (l+y \sin(\theta_{\text{in}}))}}{l+y\sin(\theta_{\text{in}})}
\approx\frac{c e^{j k (l+y \sin(\theta_{\text{in}}))}}{l},
\end{align}
where $\theta_{\text{in}}$ is the incident angle, $c$ represents the magnitude of the disturbance of the electric field at the far field point source, $l$ denotes the distance between the source and the center of the RIS, and the approximation holds because $l \gg y$. The energy flux density at $(0,0,0)$ can be calculated from electromagnetic theory or the Friis Transmission Equation \cite{3} as

\begin{equation}
D_0=\frac{\Vert \mathbf{E}(0,0) \Vert^2}{2\eta}=\frac{c^2}{2\eta l^2}
=\frac{P_t G_t}{4\pi l^2},
\end{equation}
where $P_t$ is the transmit power, $G_t$ is the transmit antenna gain. Thus, we derive the following property:
\begin{equation}
c^2=\frac{P_t G_t \eta}{2\pi}.
\end{equation}
\textcolor{black}{Define $E_r(x,y,m)$ as the electric field intensity at the $m$th antenna that is excited by a unit area of the RIS located at $(x, y, 0)$. }
Consider $(x,y,0)$, $x\in (-0.5b,0.5b)$, $y\in (-0.5a,0.5a)$ as new sources of spherical wavelets, and $E_r(x,y,m)$ can be derived from Fresnel-Kirchoff's diffraction formula as
\begin{align}
E_r(x,y,m)=\frac{c}{j\lambda}
\frac{ e^{j k (l+y \sin(\theta_{\text{in}}))}}{l}
\frac{ e^{j k d(x,y,m)}}{d(x,y,m)} \times \frac{\cos(\theta_{\text{in}})+\cos(\theta_{\text{out}}(x,y,M))}{2},
\end{align}
where $\theta_{\text{out}}(x,y,m)$ denotes the angle between $\mathbf{e}_z$ and the vector from $(x,y,0)$ to the $m$th receiver antenna, and $d(x,y,m)$ denotes the distance from $(x,y,0)$ to the $m$th receiver antenna. Define $\mathbf{E}_r=[E_r(1),\cdots,E_r(M)]^T$. We rewrite (6) as
\begin{align}
\mathbf{E}_r(x,y)&=c \mathbf{q}(x,y) \circ \mathbf{b}(x,y)\\
\mathbf{q}(x,y)&=\frac{1}{2j l \lambda}
\left[\frac{\cos(\theta_{\text{in}})+\cos(\theta_{\text{out}}(x,y,1))}{d(x,y,1)} \right.
\left.\cdots,\frac{\cos(\theta_{\text{in}})+\cos(\theta_{\text{out}}(x,y,M))}{d(x,y,M)} \right]^T \\
\mathbf{b}(x,y)&=[e^{-jk(l+y\sin(\theta_{\text{in}})+d(x,y,1))},
\left.\cdots,e^{-jk(l+y\sin(\theta_{\text{in}})+d(x,y,M))}\right]^T,
\end{align}
where $\mathbf{q}(x,y) \in \mathbb{C}^M$ denotes the scalar-multiplication of the path gain from the transmit antenna to $(x,y,0)$  and the path gain from $(x,y,0)$ to each antenna of the receiver, whereas $\mathbf{b}(x,y) \in \mathbb{C}^M$  represents the steering vector from $(x,y,0)$ to the receiver. 
Suppose the power reflected by $(x,y,0)$ and received  by the $m$th antenna forms the vector $\mathbf{p}_r(x,y) \in \mathbb{C}^M$. The channel gain per unit area is defined as $\mathbf{g}(x,y) \in \mathbb{C}^M$ formulated as
\begin{align}
\mathbf{g(x,y)}&=\sqrt{\frac{\mathbf{p}_r(x,y)}{P_t}}
=\sqrt{\frac{\left|\mathbf{E}_r(x,y)\right|^2 G_r \lambda^2}{2\eta P_t 4\pi }}
=\sqrt{\frac{c^2 G_r \lambda^2 \left|\mathbf{q}(x,y)\right|^2 \circ \left|\mathbf{b}(x,y)\right|^2 }{8\eta \pi P_t}\nonumber}\\
&\overset{(a)}{=}\sqrt{\frac{G_t G_r \lambda^2 \left|\mathbf{q}(x,y)\right|^2 }{16 \pi^2}}
=\frac{\lambda \mathbf{q}(x,y)}{4 \pi}\sqrt{G_t G_r},
\end{align}
where $\left|\cdot\right|^2$, $\sqrt{\cdot}$, and $(\cdot)^2$ are element-wise operations for vectors, $\overset{(a)}{=}$ holds because $\left|\mathbf{b}(x,y)\right|^2=[1,\cdots,1]^T$, and $G_r$ is the antenna gain of each receiver antenna.
Moreover, $\mathbf{h}$ is computed by merging the shift of phase and amplitude on the RIS, the channel gain, and the steering vector together as
\begin{align}
\mathbf{h}=\iint_{S} \tau(x,y) e^{j\beta(x,y)}\mathbf{g}(x,y) \circ \mathbf{b}(x,y)dxdy,
\end{align}
where  $S$ represents the reflecting surface of the RIS.

The goal is to design reflection coefficient $\tau(x,y)$ and $\beta(x,y)$ that can maximize $\|\mathbf{h}\|^2$ such that the received signal has the largest power \cite{article}.

\subsubsection{When The Receiver Is Equipped With a single antenna}
 Since both $\mathbf{g}(x,y)$ and $\mathbf{b}(x,y)$ are only relevant to the coordinates of each antenna and are irrelevant to the arrangement mode of the antenna array, equations (2)-(11) also hold for any other kind of multi-antenna array receiver and the single antenna receiver. For the single antenna receiver, we have $M=1$, where $\mathbf{h}$, $\mathbf{g}(x,y)$, and $\mathbf{b}(x,y)$ all degrade into a complex number.

\subsection{The Received Power}
Denote the electric field at the $m$th antenna caused by the reflected wave as $\mathbf{E}_{\text{out},m}$. Since the value of energy flux density is $D_1=\frac{\Vert \mathbf{E}_{\text{out}} \Vert^2}{2\eta}$, the received signal power $P$ can be derived by the Friis Transmission Equation \cite{3} as

\begin{align}
P=\Vert \mathbf{h} s\Vert^2=\sum_{m=1}^M \frac{\Vert \mathbf{E}_{\text{out},m} \Vert^2}{2\eta}
\frac{\lambda^2 G_r}{4\pi},
\end{align}
where $G_r$ is the antenna gain. The identical relation confirms the compatibility of channel model and electromagnetic model.

%
%

\section{Reflected Fields of the RIS}
We assume the field configuration of the incident wave is the transverse magnetic $\mathbf{e}_x$, \textit{i.e.}, the E-field is parallel to $\mathbf{e}_x$ while the H-field lies in the plane spanned by $\mathbf{e}_y$ and $\mathbf{e}_z$. Since the source is in the far field, the impinging wave field is approximated as a plane that has the electric and the magnetic field distributions:
\begin{align}
\mathbf{E}_{\text{in}} &=E_{0} e^{-j k\left(\sin \left(\theta_{\text{in}}\right) y-\cos \left(\theta_{\text{in}}\right) z\right)} \boldsymbol{e}_{x}, \\
\mathbf{H}_{\text{in}} &=-\frac{E_{0}}{\eta}\left(\cos \left(\theta_{\text{in}}\right) \boldsymbol{e}_{y}+\sin \left(\theta_{\text{in}}\right) \boldsymbol{e}_{z}\right) e^{-j k\left(\sin \left(\theta_{\text{in}}\right) y-\cos \left(\theta_{\text{in}}\right) z\right)},
\end{align}
where $\eta$ is the characteristic impedance of the medium.

According to the reflected theory \cite{textbook},  $\mathbf{E}_{\text{out}}$ and  $\mathbf{H}_{\text{out}}$ satisfy the relations
\begin{align}
\left.\mathbf{E}_{\text{out}}\right|_{z=0}&=\left.\Gamma(x, y) \mathbf{E}_{\text{in}}\right|_{z=0} \\
\mathbf{e}_{z} \times\left.\mathbf{H}_{\text{out}}\right|_{z=0}&=-\Gamma(x, y) \mathbf{e}_{z} \times\left.\mathbf{H}_{\text{in}}\right|_{z=0}.
\end{align}
Since the RIS is a large reflecting surface, the thickness of the RIS is always negligible compared to its length and width.
We suppose that the RIS can be replaced by an imaginary
surface, and the transmitted fields below the imaginary
surface are expressed by $\mathbf{E}_{\text{tr}}$ and $\mathbf{H}_{\text{tr}}$, respectively. According to
the induction theorem \cite{textbook}, $\mathbf{E}_{\text{in}}$ and $\mathbf{H}_{\text{in}}$ above
the imaginary surface can be removed. Then, an equivalent
electric current density $\mathbf{J}_{e}$ and a magnetic current density
$\mathbf{M}_{e}$ must be imposed on the imaginary surface to satisfy the boundary
conditions \cite{textbook}, which can be separately expressed as
\begin{align}
\mathbf{J}_{e} &=\mathbf{e}_{z} \times\left(\left.\mathbf{H}_{\text{out}}\right|_{z=0}-\left.\mathbf{H}_{\text{tr}}\right|_{z=0}\right), \\
\mathbf{M}_{e} &=-\mathbf{e}_{z} \times\left(\left.\mathbf{E}_{\text{out}}\right|_{z=0}-\left.\mathbf{E}_{\text{tr}}\right|_{z=0}\right).
\end{align}
\begin{figure*}[t]
\begin{minipage}[t]{0.32\linewidth}
\subfigure[planar wave]{
\includegraphics[width=5.5cm]{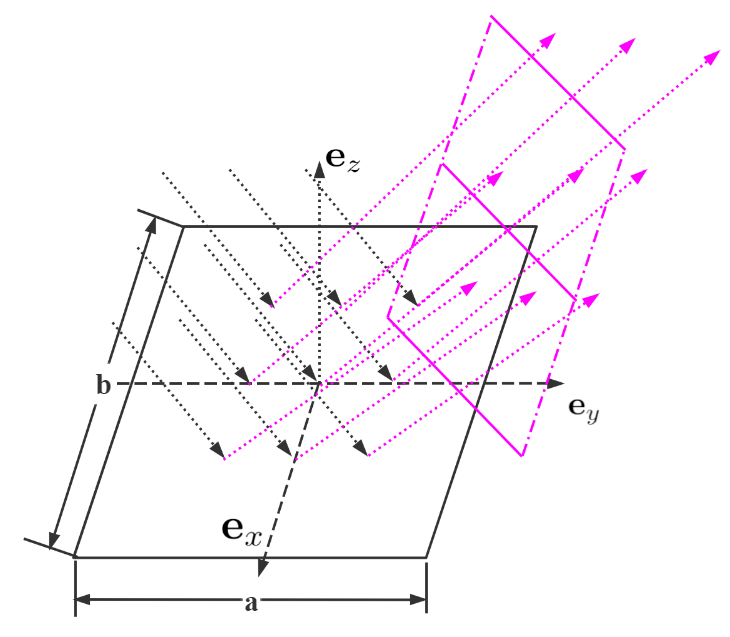}}
\end{minipage}
\begin{minipage}[t]{0.32\linewidth}
\subfigure[cylindrical wave]{
\includegraphics[width=5.5cm]{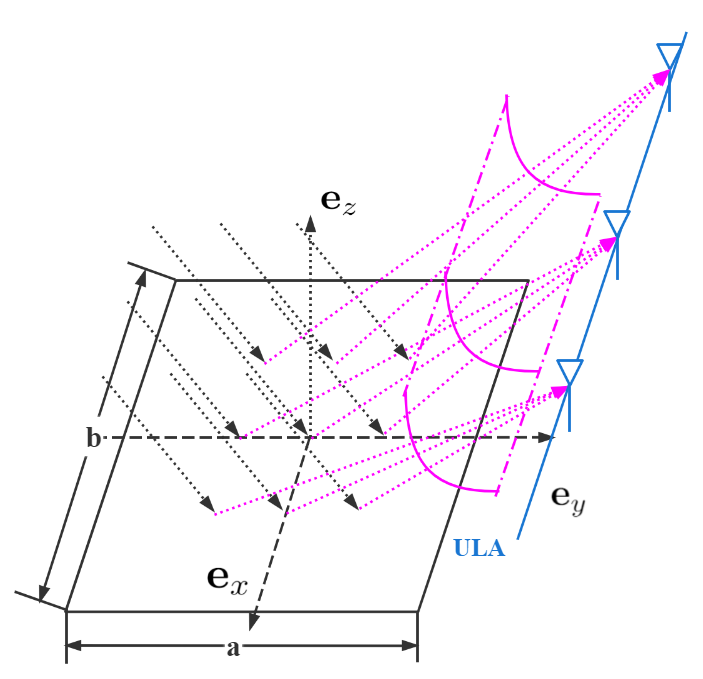}}
\end{minipage}
\begin{minipage}[t]{0.32\linewidth}
\subfigure[spherical wave]{
\includegraphics[width=6.5cm]{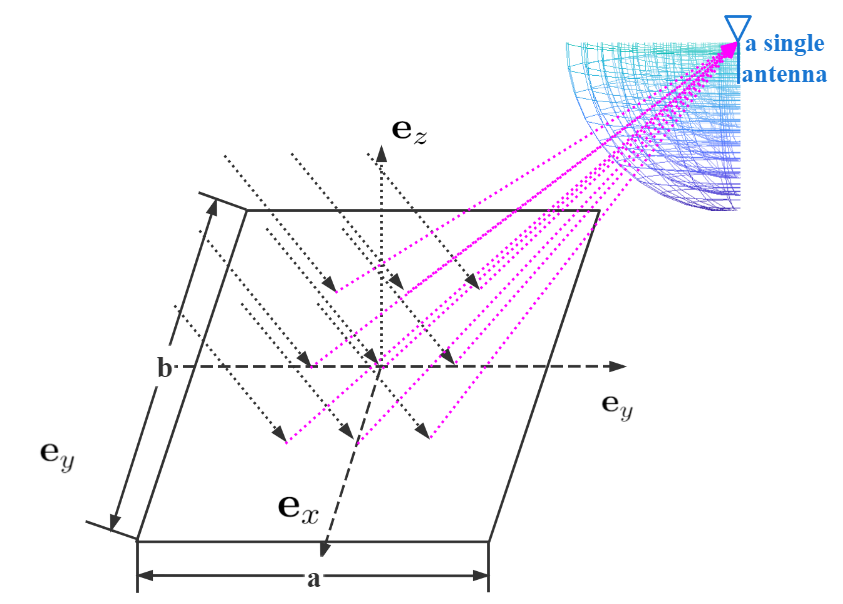}}
\end{minipage}
\caption{Comparison of Planar Reflected Wave, Cylindrical Reflected Wave, and Spherical Reflected Wave
}
\end{figure*}
Conventionally, the RIS can be assumed as a perfect magnetic conducting (PMC) surface \cite{6}, where $\mathbf{E}_{\text{tr}}$, $\mathbf{H}_{\text{tr}}$, and $\mathbf{M}_{e}$ become zero.
Hence, only $\mathbf{J}_{e}$ contributes to the scattered fields. Finally, the image theory \cite{6} is applied to remove the PMC surface and to derive an unbounded environment \cite{textbook}. The equivalent electric current density $\mathbf{J}_{f}$ is expressed as 
\begin{align}
\mathbf{J}_{f}=2 \mathbf{J}_{e}=-2 \tau e^{\beta(x, y)} \mathbf{e}_{z} \times\left.\mathbf{H}_{\text{in}}\right|_{z=0}
=-2\frac{E_0}{\eta} \tau e^{\beta(x, y)} \cos \theta_{\text{in}} e^{-j k \sin \left(\theta_{\text{in}}\right) y} \mathbf{e}_{x}=J_{x} \mathbf{e}_{x}. \hfill
\end{align}

With $\mathbf{J}_{f}$, we can compute the vector potential
$\mathbf{A}$ \cite{textbook} at an arbitrary observation point $(x^{\prime} , y^{\prime} , z^{\prime})$ as
\begin{align}
\mathbf{A}=\frac{\mu}{4 \pi} \iint_{\mathcal{S}} \mathbf{J}_{f} \frac{e^{-j k R}}{R} d x d y,
\end{align}
where $R=\sqrt{\left(x^{\prime}-x\right)^{2}+\left(y^{\prime}-y\right)^{2}+\left(z^{\prime}\right)^{2}}$ is the distance
from point $(x, y, 0)$ to this observation point. Then, $\mathbf{E}_{\text{out}}$ can be derived as 
\begin{align}
\mathbf{E}_{\text{out}}=\frac{1}{j k \sqrt{\mu \varepsilon}} \nabla \times(\nabla \times \mathbf{A})=\frac{1}{j k \sqrt{\mu \varepsilon}} \left(\nabla(\nabla \cdot \mathbf{A})-\nabla^2\mathbf{A}\right).
\end{align}
Note that (12) is only applicable for the scattered fields above the $xy$ plane.

%
%

\section{Reflection Coefficient Design Criterion of the RIS}

\subsection{Conventional Reflection Coefficient Design For Planar Wave}
\textcolor{black}{Since  $\theta_{\text{out}}(x,y)$ is identical on the whole RIS for planar waves, we use $\theta_{\text{out}}$ for simplicity.}
Conventionally, the RIS is designed to reflect the incident plane wave
as the planar wave with an identical reflected angle $\theta_{\text{out}}$ as shown in Fig. 1(a), where the following field distribution holds
\begin{align}
\mathbf{E}_{\text{out}} =E_{\text{out}} e^{-j k\left(\sin \left(\theta_{\text{out}}\right) y+\cos \left(\theta_{\text{out}}\right) z\right)} \boldsymbol{e}_{x}. 
\end{align}
Suppose the center of the receiver is located at $(0,F_y,F_z)$. In order to concentrate the reflected signal power on the receiver, one should design $\theta_{\text{out}}=\arctan(F_y/F_z)$.
With $\mathbf{E}_{\text{out}}$, $\beta(x,y)_{planar}$ can be derived from the generalized Snell’s law \cite{snell} as
\begin{align}
\beta(x,y)_{planar}&=\angle\left(\frac{\left.\mathbf{E}_{\text{out}}\right|_{z=0}\cdot \boldsymbol{e}_{x}}{\left.\mathbf{E}_{\text{in}}\right|_{z=0}\cdot \boldsymbol{e}_{x}}\right)=ky(\sin \left(\theta_{\text{in}}\right)-\sin \left(\theta_{\text{out}}\right)).
\end{align}
To provide a vivid picture on how the refection coefficient vary with the coordinates on the RIS, we present $\beta(x,y)_{\text{planar}}$ in Fig.~2.
Since the amplitudes of the incident waves and the reflected waves are identical on each point of the RIS, the amplitudes of the reflection coefficient are  equal to $1$, \textit{i.e.},
\begin{align}
\tau(x,y)_{planar}=1.
\end{align}
Note that, the conventional communication focuses on the direction of beamforming but does not pay attention to the propagation of electromagnetic waves \cite{9398864}. However, the planar wave can not focus the incident energy on ULA or a single antenna in the near field, which will be discussed from EM theory in the next subsection. 

\subsection{Reflection Coefficient Design For the ULA receiver}
Consider a ULA receiver being located in the  near field of the RIS, which may be caused by the large scale of the RIS, the short distance of the receiver, or the high frequency of electromagnetic waves according to (1).
Since the LOS path from each points of the RIS and the receiver can not be regarded parallel, it is inappropriate to adopt the conventional identical reflection angle scheme. In order to create a focal line in the position of the ULA receiver, we need to design a more sophisticated reflection coefficient scheme to realize variable reflection angles.
Suppose the ULA receiver is located at $(0,F_y,F_z)$ and is perpendicular to the $yz$ plane.
In order to focus the incident energy, the optimal scattered waves should converge on the focal line. To achieve the constructive interference, all the waves should share the same phase on the focal line. Thus, when we consider the time reverse of the optimal waves (the backward propagation of the optimal waves that is a standard operation in \cite{PhysRevX.6.041008}), the waves can be regarded as radiated by the same source of radiation with the current intensity defined as $I_1$ on the focal line. Suppose the radiation distribution is consistent with the change of $x$, and then the source of radiation can be seen as a line source parallel to $\boldsymbol{e}_{x}$ located at $(0,F_y,F_z)$ with infinite length. Since the time reverse of the optimal scattered waves are axially symmetric with the focal line as the axis, the optimal scattered waves share the same property, \textit{i.e.},
the optimal scattered waves for ULA should have cylindrical wavefronts as shown in Fig. 1(b),
from which the scattered electric field $\mathbf{E}_{\text{out}}$ and magnetic field $\mathbf{H}_{\text{out}}$ can be expressed as \cite{textbook}
\begin{align}
\mathbf{E}_{\text{out}}=& \left(\frac{-I_{1} k \eta}{4} H_{0}^{(2)}\left(k R_c\right)\right)^* \boldsymbol{e}_{x}, \\
\mathbf{H}_{\text{out}}=& \left(\frac{-j I_{1} k}{4} H_{1}^{(2)}\left(k R_c\right)\right)^* \!\!
\times\!\!\left(\frac{(z-F_z)
\boldsymbol{e}_{y}}{R_c}-\frac{(y-F_y)\boldsymbol{e}_{z}}{R_c}\!\right),
\end{align}
where $R_c=\sqrt{(y-F_y)^{2}+(z-F_z)^{2}}$, $I_{1} \in \mathbb{C}$ denotes the current intensity of the line source, and $H_{n}^{(2)}$ refers to the Hankel function of type $2$ with order $n$. Without loss of generality, $\angle I_1$ is assumed to be $0$.

With $\mathbf{E}_{\text{out}}$, $\beta(x,y)_{\text{cylind}}$ can be derived as
\begin{align}
\beta(x,y)_{\text{cylind}}&=\angle\left(\frac{\left.\mathbf{E}_{\text{out}}\right|_{z=0}\cdot \boldsymbol{e}_{x}}{\left.\mathbf{E}_{\text{in}}\right|_{z=0}\cdot \boldsymbol{e}_{x}}\right)
\!\!=\!\!\angle\left(\frac{(\frac{-I_{1} k \eta}{4} H_{0}^{(2)}\left(k R_c\right))^*}{E_0 e^{-j k\left(\sin \left(\theta_{\text{in}}\right) y\right)}}\right) \hfill \nonumber\\
&=-\angle\left(-H_{0}^{(2)}\left(k R_c\right)\right)+k\sin \left(\theta_{\text{in}}\right)y.
\end{align}

To provide a vivid picture on how the refection coefficient vary with the coordinates on the RIS, we calculate $\beta(x,y)_{\text{cylind}}$ when $a=20\lambda$ and the incident angle is $\pi/6$, as an example in Fig.~3, where the focal line is parallel to $\mathbf{e}_x$ such that $\beta(x,y)_{\text{cylind}}$ is irrelevant to $x$ and is only relevant to $y$. The tangent slope of $\beta(x,y)_{\text{cylind}}$ changes from positive to negative with the reflection angle decreases from $y=-10\lambda$ to $y=10\lambda$. When the tangent slope of $\beta(x,y)_{\text{cylind}}$ is zero, it indicates that at this conventional-mirror-reflection point the reflection angle is equal to the incident angle like conventional mirror reflection. 
With the decrease of $y$, $\beta(x,y)_{\text{cylind}}$ is increasingly analogous to a linear function, as the reflection angle changes less obviously.

\begin{figure}[t]
  \centering
  \centerline{\includegraphics[width=10.7cm]{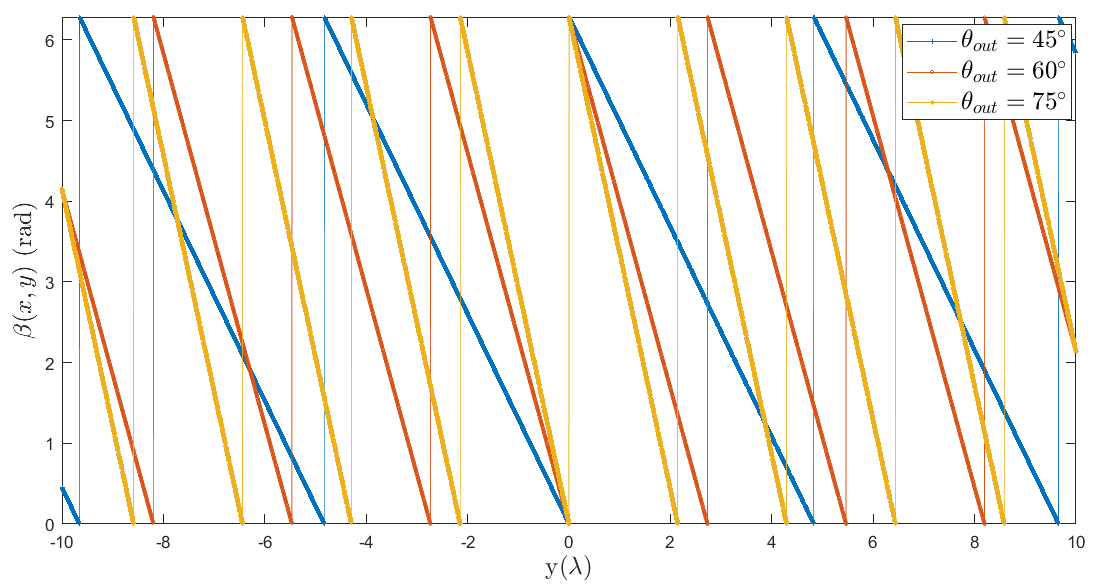}}
  \caption{Phase shift on the RIS  to convert the planar wave into the planar wave. In this case, $\beta(x,y)$ is irrelevant to $x$.}
\end{figure}
When the reflected waves are cylindrical waves, the amplitudes of the scattered radiation on each point of the RIS rely on $R_c$ and $|I_{1}|$. According to the location of the focal line, $R_c$ is known and $|I_{1}|$ can be calculated by $E_0$.
With the assumption that the RIS is passive lossless, the power conservation determines the relation between $E_0$ and $|I_{1}|$. The power of the incident wave on the RIS and the power of the reflected wave on the RIS  are
\begin{align}
P_{\text {incident}} &=\frac{E_{0}^{2}}{2 \eta} a b \cos(\theta_{\text{in}}), \\
P_{\text {reflected}} &=\frac{\tan ^{-1}((F_y + 0.5a)/ F_z)-\tan ^{-1}((F_y - 0.5a)/ F_z)}{2\pi}\times \frac{|I_{1}|{ }^{2} k \eta b}{16 \pi},
\end{align}
respectively. When $P_{\text {incident}} = P_{\text {reflected}}$, $|I_{1}|$ is determined by $E_0$, and the desired amplitude of the reflection coefficient is derived by the definition, as
\begin{align}
\tau(x,y)_{\text{cylind}}=\left|\left(\frac{(\frac{-|I_{1}| k \eta}{4} H_{0}^{(2)}\left(k R_c\right))^*}{E_0 e^{-j k\left(\sin \left(\theta_{\text{in}}\right) y\right)}}\right)\right|. \hfill 
\end{align}

\begin{figure}[t]
  \centering
  \centerline{\includegraphics[width=10.7cm,height=8cm]{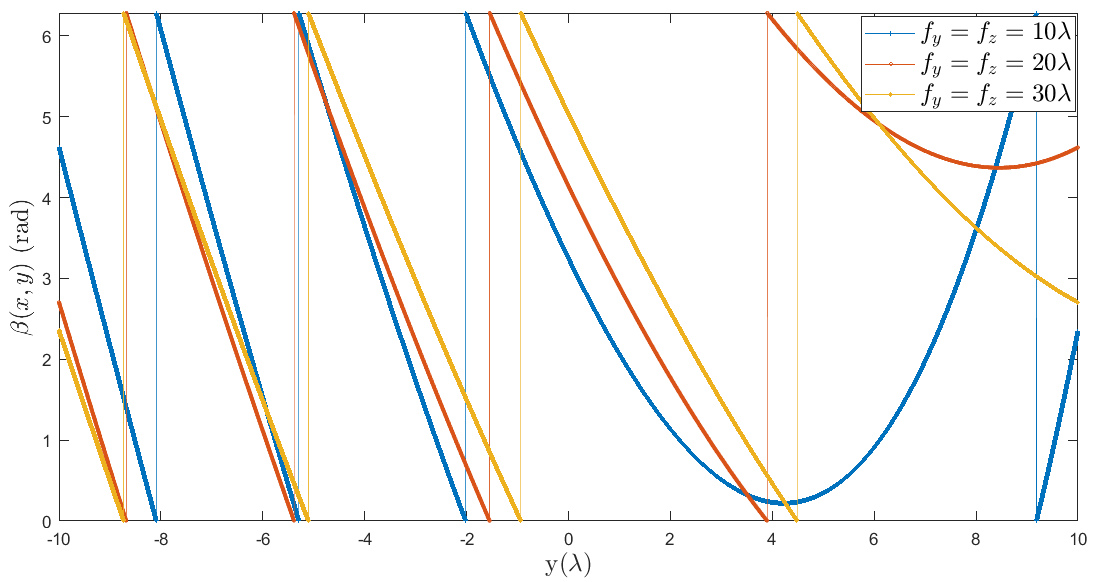}}
  \caption{Phase shift on the RIS  to convert the planar wave into the cylindrical wave. In this case, $\beta(x,y)$ is irrelevant to $x$.}
\end{figure}

\begin{figure}[t]
  \centering
  \centerline{\includegraphics[width=10.7cm,height=8cm]{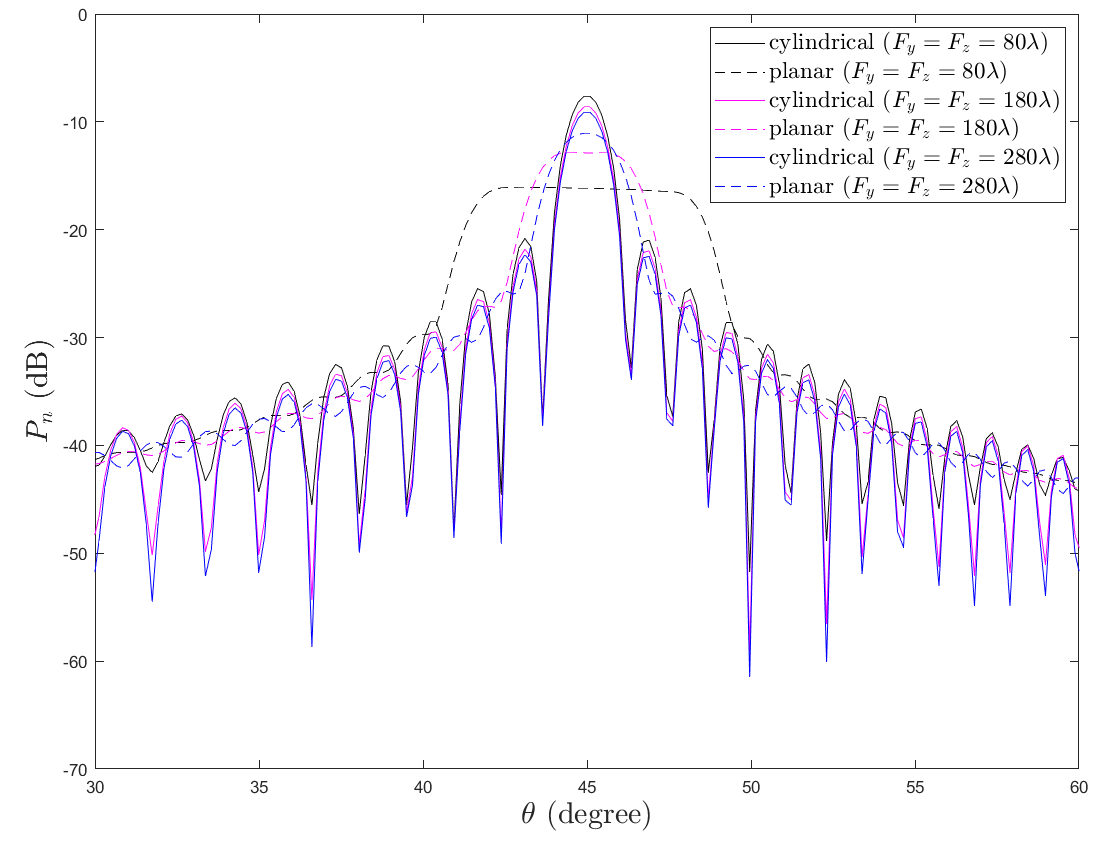}}
  \caption{The comparison of normalized power between planar wave and cylindrical wave}
  \label{能量分布比较}
\end{figure}
For the ULA receiver, the power should be concentrated on the focal line, which cannot be achieved by traditional planar waves. To demonstrate the superiority of  cylindrical waves over planar waves, we examine the degree of energy leakage, \textit{i.e.}, the energy carried by the reflected waves to the undesired area when $a = b = 20 \lambda$ and the incident angle is $\pi$/6. We select the observation points at the circular arc of $x=0$, $y=d\cos(\theta)$, $z=d\sin(\theta)$, $\theta \in (0\degree, 90\degree)$ and calculate the electric field $\mathbf{E}_{\text{obs}}$ on the observation points. 
\textcolor{black}{To show the change of the focusing property when the ULA moves from small distance to large distance, three groups of focal points  are set with their location in the order as $(80\lambda, 80\lambda)$, $(180\lambda, 180\lambda)$, and $(180\lambda, 180\lambda)$ respectively. } The distribution of the normalized power $P_{n}=\frac{\Vert \mathbf{E}_{\text{obs}} \Vert^2}{2\eta d^2}$ versus $\theta$ is shown in Fig.~\ref{能量分布比较}.
It is seen that, for the cylindrical wave, the power is the maximum at $\theta=45\degree=\theta_{F}$ that is precisely the angular position of the ULA, and the majority of the power is concentrated around the focal line. The degree of power concentration keeps almost unchanged when the focal line moves closer to the center of the RIS.
For the planar wave, we design the reflection coefficient such that the reflection angle is equal to $\theta_{F} =45\degree$, which is the optimal reflected angle to concentrate power on the ULA.
When $F_y=F_z=80\lambda$, the main lobe width is significantly larger than the proposed cylindrical wave and the maximum power is 8.53 dB smaller than the proposed cylindrical wave, which results in serious energy leakage.
As the focal line moves farther to the center of the RIS, the degree of power concentration increases and the power distribution of the planar wave is increasingly similar to the power distribution of the cylindrical wave.

\subsection{Reflection Coefficient Design For A Single Antenna Receiver}

Consider the receiver is equipped with a single antenna whose location $(F_x,F_y,F_z)$ is the focal point of the RIS as shown in Fig. 1(c).
In order to focus the incident energy on this focal point, the optimal scattered waves should converge on the focal point. To achieve constructive interference, all the waves should share the same phase at the focal point. Thus, when we consider the time reverse of the optimal waves (the backward propagation of the optimal waves), they can be regarded as being radiated by the same point source with the magnitude of radiation $U_1$ at the focal point. Therefore, the optimal reflected waves for the single antenna should have spherical wave fronts, and 
the amplitude of the scattered electric field $E_{\text{out}}$ can be expressed as \cite{textbook}
\begin{align}
E_{\text{out}}= \frac{U_1 e^{jkR_s}}{R_s},
\end{align}
where $R_s=\sqrt{(x-F_x)^{2}+(y-F_y)^{2}+(z-F_z)^{2}}$, and $\angle\left(U_1\right)$ can be regarded as $0$ without loss of generality.
We note that the above field distribution is
accurate if an ideal drain, \textit{i.e.}, the time-reversed equivalent
of the point source, is also positioned in the focal point. In
practice, the absence of the ideal drain may reduce the overall
performance of the designed scheme.
With $\mathbf{E}_{\text{out}}$, $\beta(x,y)_{\text{sphere}}$ is derived as 
\begin{align}
\beta(x,y)_{\text{sphere}}=\angle\left(\frac{E_{\text{out}}}{\left\Vert\mathbf{E}_{\text{in}}\right\Vert}\right)
=-k R_s+k\sin \left(\theta_{\text{in}}\right)y.
\end{align}
\begin{figure}[t]
  \centering
  \centerline{\includegraphics[width=10.7cm,height=8cm]{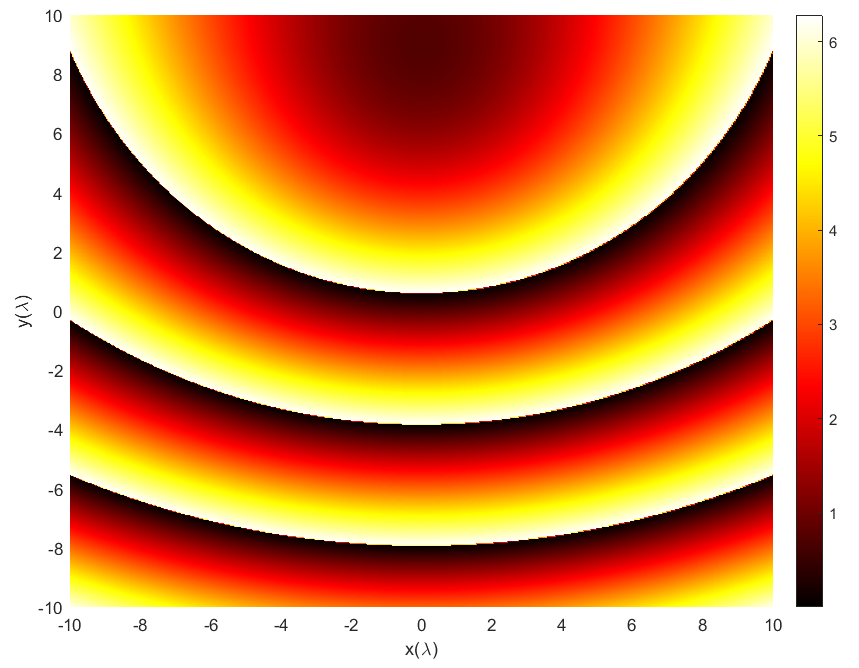}}
  \caption{Phase shift on the RIS is designed to convert the planar wave into the spherical wave.}
  \label{球面波相位图}
\end{figure}
To provide a vivid picture on how the refection coefficient vary with the coordinates on the RIS, we present $\beta(x,y)_{\text{sphere}}$ that is a hyperboloidal function plus a linear function, as an example in Fig.~\ref{球面波相位图}, when $(F_x,F_y,F_z)=(0,0,20\lambda)$ and the incident angle is $\pi/6$.

While the phase shift is essential to focus the radiated field on the focal point, the amplitude shift can be used to reduce 
the sidelobe level. Indeed, a high level of the secondary lobes around the focal point may degrade measurement accuracy in 
non-contact sensing applications. Meanwhile, high sidelobes may reduce 
transmission efficiency in wireless power transfer systems, while increase the interference to nearby wireless systems.
In near-field-focused microwave antennas \cite{7912361}, \cite{6948345}, it has been observed 
that an amplitude shift that gives lower transverse sidelobes also yields higher forelobes and aftlobes along the aperture axis.
The amplitudes of the scattered radiation on each point of the RIS rely on $R_s$ and $|U_1|$. According to the location of the focal point, $R_s$ is known and $|U_{1}|$ can be calculated by $E_0$.
With the assumption that the RIS is passive lossless, the power conservation determines the relation between $E_0$ and $|U_{1}|$. The power of the incident wave on the RIS and the power of the reflected wave on the RIS  are
\begin{align}
P_{\text {incident}} &=\left|\frac{1}{2}\left(\mathbf{H}_{\text{in}}\times\mathbf{B}_{\text{in}}\right)\cdot\mathbf{S}\right|=\frac{E_{0}^{2}}{2 \eta} a b \cos(\theta_{\text{in}}), \\
P_{\text {reflected}} &=\Omega \frac{|U_{1}|^2}{2\eta},
\end{align}
where $\mathbf{S}$ denotes the normal direction vector of $S$, and $\Omega$ denotes the solid angle between the RIS and the focal point. When $P_{\text {incident}} = P_{\text {reflected}}$, $|U_{1}|$ is determined by $E_0$, and the desired amplitude of the reflection coefficient is derived by the definition as
\begin{align}
\tau(x,y)_{\text{sphere}}=\left|\left(\frac{|U_{1}|}{R_s E_0 e^{-j k\left(\sin \left(\theta_{\text{in}}\right) y\right)}}\right)\right|. \hfill 
\end{align}
Note that the  spherical wave applies to more than the single receiver model. If the ULA receiver is relatively short or is composed of relatively few antenna, then the ULA degenerates into a point. In this case, the RIS reflection coefficient design is the same as (31)-(35).

\begin{figure}[t]
  \centering
  \centerline{\includegraphics[width=10.7cm,height=8cm]{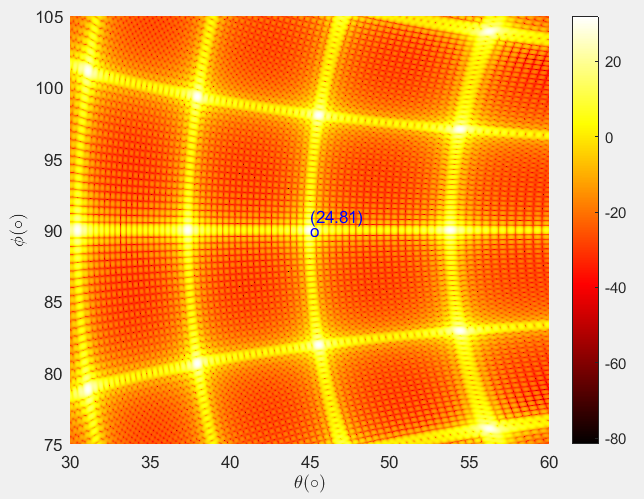}}
  \caption{The normalized power (dB) of spherical wave for single antenna}
  \label{power_spherical}
\end{figure}
\begin{figure}[t]
  \centering
  \centerline{\includegraphics[width=10.7cm,height=8cm]{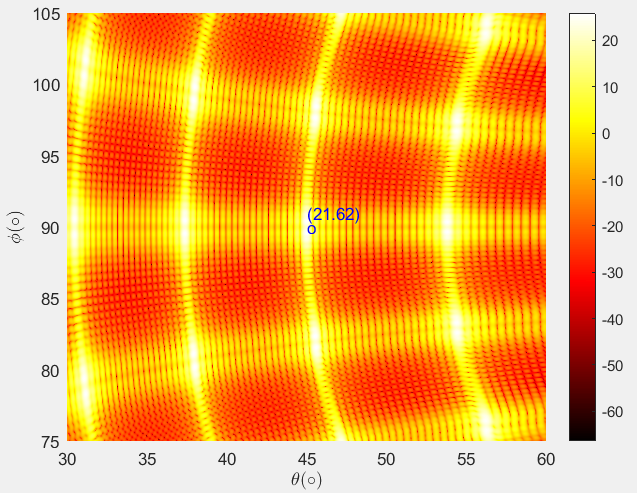}}
  \caption{The normalized power (dB) of cylindrical wave for single antenna}
  \label{power_cylindrical}
\end{figure}
In order to demonstrate the distribution of the power in the whole space, we provide the following example. We set $L=2m$, $\lambda=0.01m$, and adjust the transmit power such that the normalized power at the focal point is 1 dB when planar waves are reflected.
We set the single antenna as the focal point for the spherical waves at $d_0=50m$, $\phi_s =90\degree$, $\theta_s =45\degree$. For the cylindrical wave, since the focusing property is the best when the single antenna is on the focal line, we set the focal line being perpendicular to the $yz$ plane with its center at $d_0=50m$, $\theta_0 =45\degree$. Then, we calculate the electric field $\mathbf{E}_{\text{obs}}$ on the observation points for cylindrical waves and spherical waves respectively.
\textcolor{black}{
In order to show the superiority of the spherical waves in the case of single antenna, we select the observation points on the spherical surface where $d=50m$, $\phi_s \in (0\degree,180\degree)$, $\theta_s \in (0\degree, 90\degree)$, while $\phi_s$ is the azimuth angle and $\theta_s$ is the elevation angle. }
The distributions of the normalized power $P_{n}=\frac{\Vert \mathbf{E}_{\text{obs}} \Vert^2}{2\eta d^2}$ for spherical waves and for cylindrical waves are shown in Fig.~\ref{power_spherical}.
It is seen that for the cylindrical wave, the power is the maximum on the line of $\theta_s = 45\degree$, $\phi_s \in (0\degree,180\degree)$, and the majority of the power is concentrated around $\phi_s \in (80\degree,100\degree)$ on the focal line. 
Since the size of the RIS is limited, the ability of cylindrical wave to focus the power on the focal line gradually decreases when $\phi_s$ moves away from $90\degree$.
For the spherical wave, the power is maximum at the point of $\theta_s = 45\degree$, $\phi_s =90\degree$, and the majority of the power is concentrated within $\theta_s \in (40\degree,50\degree)$ around the focal point. 
The power decreases sharply when the distance between the observed point and the focal point increases. Since the power is concentrated in a smaller area than the cylindrical wave and the maximum power is 3.19 dB larger than the cylindrical wave in Fig.~\ref{power_cylindrical}, the spherical wave outperforms the cylindrical wave in the capability of focusing power on a single antenna.


\section{Sensing The Location of The Receiver}
Different from far field beamforming  that only needs the knowledge of $\mathbf{h}$ or the direction of the receiver, RIS coefficient design for the near field transmission needs the exact location information of the receiver. Hence, the near field transmission actually belongs to location-awared technologies \cite{5230781}.  

We assume that the source in the far field is fixed (typically the BS) and the receiver in the near field is mobile with uncertain location. Thus, the incident angle $\theta_{in}$ can be assumed as a known constant, while the location  of the mobile receiver is unknown.

In general, there are three ways to obtain the the position information of the mobile receiver.
\begin{enumerate}
\item The first way is similar to the conventional channel estimation based beamforming, that is, the optimal beamforming coefficient is obtained by calculating the channel and then matching $\mathbf{h}$.
\textcolor{black}{
However, the instantaneous channel is difficult to estimate due to the acquisition delay and the feedback overhead.
Moreover, since the RIS here has continuous aperture,  even if  $\mathbf{h}$ can be estimated, we cannot directly obtain the reflection coefficient of the RIS at the receiver.  }
An alternative is to extract the position of the receiver through formula (11), and then use these two parameters to calculate the optimal coefficients from (27), (30), (32), and (35).  This approach is similar to the traditional angular domain based  beamforming method for massive MIMO  \cite{9398864}, \cite{8753608}, \cite{8333702}.  However, for continuous RIS, it is quite difficult to obtain the physical parameters of receiver by the inverse integration from formula (11).  
\item The second way may refer to the sensors to obtain the position  information of the physical object \cite{9705498,9705498,9687468}. For example, one can use the passive sensors like camera to  obtain near field environmental information around RIS, and then use computer vision technology to capture position information of the receiver from the image \cite{9580233,9129762,9512383}.  Another advantage of camera is that it does not need radio frequency (RF) link for signal processing, which is especially suitable for RIS-assisted transmission scenes.  
\item The third way refers to the conventional beam scanning method, which searches the position of the receiver in space, and then the receiver feeds back the time slot with the strongest receiving power to the transmitter to determine the optimal beam direction.  Different from traditional far-field beam scanning that only search over the azimuth and elevation angle, the near-field scanning has to scan beam azimuth, elevation, as well as distance at the same time. Thereby the scanning here should be one dimension higher than the conventions, which is a reasonable price paid by looking into near field communications.  Nevertheless, the advantage of this approach is that the perception can be completed by communication link itself without additional equipment. 
\end{enumerate}

\textcolor{black}{
In this paper, we adopt the beam scanning method and  propose two approaches to sense the location of the receiver, i.e., maximum likelihood (ML) method and focal scanning (FS) method.
During the beam scanning process, the reflecting coefficients are adjusted $T$ times to transmit $T$ signals. Each $\tau_t(x,y)$ and $\beta_t(x,y)$
correspond to the received $t$th signal of the $m$th antenna, denoted as $y_{m, t}$.}

\textcolor{black}{
In order to achieve higher SNR at the receiver, we design the RIS coefficients by methods proposed in \uppercase\expandafter{\romannumeral4} such that the imaginary focal point (line) scans through the search area when the $T$ signals are transmitted. Thus, each signal is received when the imaginary focal point (line) is relatively close to the receiver.
Let $\Xi=(x_u,y_u,z_u)$  denote the unknown position of the ULA's center.
 The ULA  is assumed to be perpendicular to the $yz$ plane and the location of the $m$th antenna is $\Xi_m=(x_u+L(\frac{m-1}{M-1}-\frac{1}{2}),y_u,z_u)$.
}

\subsection{Sensing by The ML Method}
\textcolor{black}{
Let $a_{m,t}$ and $\phi_{m,t}$ denote the unknown amplitude and phase angle of the $t$th noise-free signal received by the $m$th antenna. The distance between the source and the center of the RIS $l$ is assumed unknown.
The log-likelihood function is given by
\begin{align}
f\left(\left\{{y_{m, t}}\right\} \mid \Xi,\{a_{m,t}\},l \right) 
=-\frac{1}{\sigma^{2}} \sum_{m=1}^{M} \sum_{t=1}^{T}\left|y_{m, t} - a_{m,t} e^{j\phi_{m,t}} \right|^{2}
\label{log-likelihood}
\end{align}
where $\phi_{m,t}$ can be expressed as 
\begin{align}
\phi_{m,t}=\angle\left(\iint_{S} \tau_t(x,y) e^{j\beta_t(x,y)}\frac{e^{-jk(y\sin(\theta_{\text{in}})+d(x,y,m))})}{d(x,y,m)}dxdy\right)-kl=\psi_t(\Xi)-kl,
\end{align}
in which we take out the common unknown phase offset $kl$ such that
$\psi_t(\Xi)$ depends only on $\Xi$ and $t$.
}

\textcolor{black}{
By adopting the ML criterion to estimate $a_{m, t}$, we have
\begin{align}
\hat{a}_{m, t}=& \arg \max _{a_{m, t}} f\left(\left\{{y_{m, t}}\right\} \mid \Xi,\{a_{m,t}\},l \right) \nonumber \\
=& \arg \max _{a_{m, t}}\left(-a_{m, t}^{2}+2 a_{m, t} \tilde{a}_{m, t} \cos \left(\tilde{\phi}_{m, t}-\psi_t(\Xi)+kl\right)\right) \nonumber \\
=& \tilde{a}_{m, t} \cos \left(\tilde{\phi}_{m, t}-\psi_t(\Xi)+kl\right),
\end{align}
where $\tilde{a}_{m, t}=\left|y_{m, t}\right|$ and $\tilde{\phi}_{m, t}=\angle \left(y_{m, t}\right)$ are the amplitude and argument of the observed signal $y_{m, t}$, respectively. Substituting $\hat{a}_{m, t}$ for $a_{m, t}$ in (\ref{log-likelihood}, we obtain 
\begin{align}
f\left(\left\{{y_{m, t}}\right\} \mid \Xi,l \right) 
&=-\frac{1}{\sigma^{2}} \sum_{m=1}^{M} \sum_{t=1}^{T}\tilde{a}_{m, t}^{2}\sin^2\left(\tilde{\phi}_{m, t}-\psi_t(\Xi)+kl\right) \\
\hat{\Xi}&=\arg\max_{\Xi,l}f\left(\left\{{y_{m, t}}\right\} \mid \Xi,l \right) .
\label{brute}
\end{align}
}

\textcolor{black}{
Let $I_s$ denote the sampling number of the continuous integral operation in (11).
If the RIS coefficients are randomly selected during the $T$ signals, then the complexity of ML algorithm is $\mathcal{O} (I_s T M N_{tot})$,
where $N_{tot}$ represents the number of test points in case that a
brute-force minimization method is used in (\ref{brute}). The
choice of $N_{tot}$ depends on the sensing resolution $\Delta_{ML}$,
the size of the search area $A$, and the number of test values $N_l$
for $l$ (Note that $f\left(\left\{{y_{m, t}}\right\} \mid \Xi,l \right)$ is a periodic function of period $\pi/k$ in regard of $l$). For 2-D space, there is $N_{tot}$ = $N_l A /\Delta_{ML}^2$; while for 3-D space, there is $N_{tot}$= $N_l A/\Delta_{ML}^3$. 
Since the RIS coefficients are actually not randomly selected, and the design of RIS coefficients to reflect each signal demands $\mathcal{O}(I_s)$ operations, the complexity of ML algorithm is  $\mathcal{O} (I_s^2 T M N_{tot})$.
}

\subsection{Sensing by The FS Method}
\textcolor{black}{
Suppose the desired resolution for the FS method is $\Delta_{FS}$, and there is $T=A /\Delta_{FS}^2$  for 2-D space and $T=A /\Delta_{FS}^3$  for 3-D space. 
Suppose $\textbf{p}(t)$ denotes the location of the $t$th imaginary focal point (line), and the estimation of $\Xi$ by the FS method is 
\begin{align}
\hat{\Xi}=\textbf{p}(\tilde{t}), \quad \tilde{t}=\arg\max_{t} \sum_{m=1}^{M} |y_{m, t}|^2.
\label{brute2}
\end{align}
When cylindrical waves are reflected for an imaginary focal line, $\textbf{p}(t)$ only contains $y$-coordinate and $z$-coordinate and the sensing is performed in 2-D space. When spherical waves are reflected for an imaginary focal point, the sensing is performed in 3-D space.}

\textcolor{black}{
Note that the sensing resolution $\Delta_{FS}$ is implicit in $T$. Since the computation complexity of adjusting the RIS coefficients to transmit a single pilot symbol is $\mathcal{O} (I_s)$, the complexity of FS algorithm is $\mathcal{O} (I_s T M)$, which is much lower than the ML algorithm.
}

\subsection{The Position Error Bound}

\textcolor{black}{
In this section, the derivation of the position error bound (PEB) on the receiver's position
estimation error is provided as a benchmark for the ML and FS methods.
The PEB can be obtained by computing the Fisher information matrix  $\mathbf{J}(\Xi)$. Suppose $q_{t}$ is the $t$-th 
transmitted signal and $s_{m, t}(\Xi)$ is the $t$-th received signal of the $m$-th antenna without noise. Then PEB is defined as \cite{PEB}
\begin{align}
\text{PEB} \triangleq \sqrt{\operatorname{trace}[\mathbf{J}^{-1}(\Xi)]},
\end{align}
where the elements of $\mathbf{J}(\Xi)$ are
\begin{align}
[\mathbf{J}(\Xi)]_{i, j}=\frac{2}{\sigma^{2}} \Re\left\{\sum_{m=1}^{M} \sum_{t=1}^{T} \frac{\partial s_{m, t}(\Xi)}{\partial \Xi_{i}} \frac{\partial s_{m, t}^{*}(\Xi)}{\partial \Xi_{j}}\right\}.
\label{FIM}
\end{align}
The first derivative in (\ref{FIM}) is given by
\begin{align}
 \frac{\partial s_{m, t}(\Xi)}{\partial \Xi_{i}} &= \frac{G_t G_r}{8\pi jl}(\cos(\theta_{\text{in}})+\cos(\theta_{\text{out}})) q_{t} \iint_{S} \tau_t(x,y) e^{j\beta_t(x,y)}\frac{e^{-jk(l+y\sin(\theta_{\text{in}})+d(x,y,M))}}{d(x,y,m)}\nonumber\\
 &\times(-\frac{1}{d(x,y,m)}-jk)    \frac{\partial d(x,y,m)}{\partial \Xi_{i}}   dxdy,
\end{align}
where 
\begin{equation}
\frac{\partial d(x,y,m)}{\partial x_u}=\frac{x_u+L(\frac{m-1}{M-1}-\frac{1}{2})-x}{d(x,y,m)},  \frac{ \partial d(x,y,m)}{\partial y_u}=\frac{y_u-y}{ d(x,y,m)},  \frac{\partial d(x,y,m)}{\partial z_u}=\frac{z_u}{d(x,y,m)}.
\end{equation}
}

\section{Simulation Results and Analysis}
In the simulations, we set $a = b =2$m and $\lambda$ = 0.01 m, \textit{i.e.}, the frequency $f=29.98$ GHz and thus the radiating near field requires $29.4923m< d_0 < 800m$. 
Note that such range of $d_0$ is typical in practice, \textit{e.g.}, in road traffic the distance from vehicles to the nearest RIS may fall in this range.
We set $E_{\text{in}} = 1$ V/m, $M=128$, $\eta=377$ Ohm, $G_t=G_r=5$dB, $L=3$m and $\theta_{\text{in}} =30\degree$. The ULA is assumed to be perpendicular to the $yz$ plane and the center of ULA is assumed in the $yz$ plane.
Although the magnitudes of electromagnetic fields are related to $E_{\text{in}}$ and $\eta$ that will change in practice, the design of the reflection coefficient is irrelevant to $E_{\text{in}}$ and $\eta$ according to (23)-(35). Moreover, the variation trend of the received power in the sensing process is irrelevant to $E_{\text{in}}$ and $\eta$ too, which means that the sensing accuracy will not be affected by the change of $E_{\text{in}}$ and $\eta$.

We define SNR as the ratio of the power at the receiver to the power of the noise, when the planar waves are reflected in the direction from the center of the RIS to the center of the receiver. Thus we can fairly compare the growing speed of the channel capacity for the proposed cylindrical and spherical waves to the conventional planar waves.
The channel capacity is calculated as 
\begin{align}
C=\log_2\left(1+\sum_{m=1}^M \frac{\Vert \mathbf{E}_{\text{out},m} \Vert^2}{2\eta}
\frac{\lambda^2 G_r}{4\pi N}\right),
\end{align}
where $N$ denotes the power of the additive noise.

\textcolor{black}{
The signals for FS and ML methods are both generated when we adjust the coefficients of the RIS such that the imaginary focal point (line) scans through the search area.
The $t$-th transmitted signal $q_{t}$ is constantly equal to $1$ during the scanning process.
The sensing resolution is set as $\Delta_1=\Delta_{ML}=\Delta_{FS}=10^{-4}d_0$ for ULA receiver, and $\Delta_2=\Delta_{ML}=\Delta_{FS}=2\times10^{-3}d_0$ for the single antenna receiver.
}
\subsection{Sensing The Location of ULA}
\begin{figure}[t]
  \centering
  \centerline{\includegraphics[width=10.7cm,height=8cm]{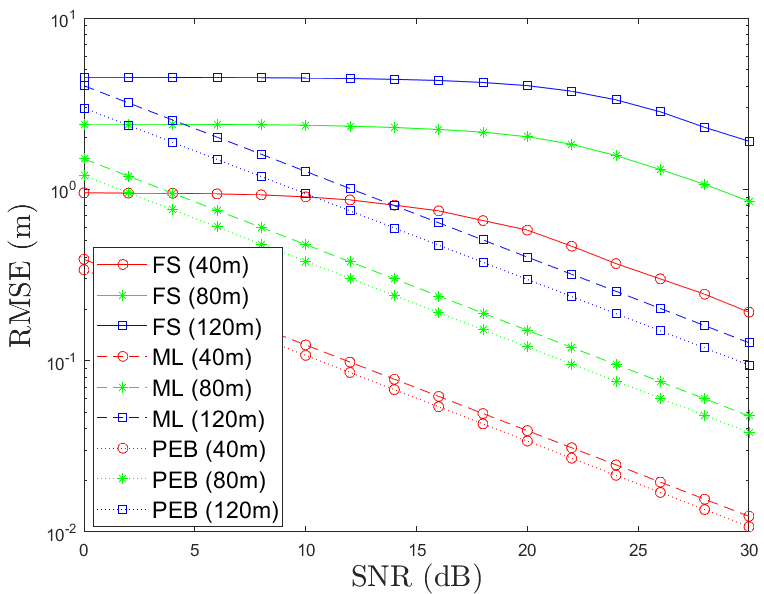}}
  \caption{The $\text{RMSE}$ of sensing location varies with SNR for the ULA.}
  \label{sensing_snr1}
\end{figure}
\textcolor{black}{
Since the center of ULA is assumed in the $yz$ plane, the sensing for ULA is performed in 2-D space.
The center of ULA is randomly located with uniform distribution
in the area $(0,\frac{d_0}{\sqrt{2}}-250\Delta_1,\frac{d_0}{\sqrt{2}}-250\Delta_1)\times(0,\frac{d_0}{\sqrt{2}}+250\Delta_1,\frac{d_0}{\sqrt{2}}+250\Delta_1)m^2$.
In Fig.~\ref{sensing_snr1}, we investigate the $\text{RMSE}$ of sensing location versus SNR with $d_0=40$m, $d_0=80$m, and $d_0=120$m respectively. Cylindrical waves are reflected for signals of both FS and ML methods. PEB is shown as the benchmark.
For all cases, the $\text{RMSE}$ decreases with the increase of SNR. 
It is seen that the smaller $d_0$ is, the more accurate the sensing will be, because the near field effect is more significant when the RIS is closer to the receiver.
The estimation error for FS is higher than ML because FS only utilizes part of the
information corresponding to the max-amplitude signal during the scanning process, while the other information
is wasted. The sensing accuracy of ML is close to PEB. However, the complexity of FS is much lower
than that of ML. Therefore, ML is applicable to scenarios where high sensing accuracy is demanded, while FS is applicable to scenarios where small computation overhead is demanded.
}



\subsection{Sensing The Location of The Single Antenna}
\begin{figure}[t]
  \centering
  \centerline{\includegraphics[width=10.7cm,height=8cm]{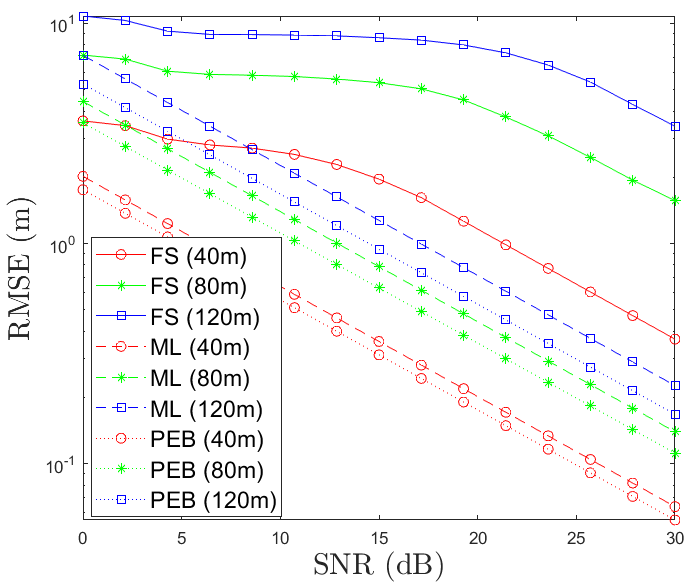}}
  \caption{The $\text{RMSE}$ of sensing location varies with SNR for the single antenna receiver.}
  \label{sensing_snr_single}
\end{figure}
\textcolor{black}{
In this simulation, we investigate the $\text{RMSE}$ of sensing location versus SNR in Fig.~\ref{sensing_snr_single} with $d_0=40$m, $d_0=80$m, and $d_0=120$m respectively. Spherical waves are reflected to scan through the search area for signals of both FS and ML methods.
The single antenna is randomly located with uniform distribution
in the area $(-150\Delta_2,\frac{d_0}{\sqrt{2}}-150\Delta_2,\frac{d_0}{\sqrt{2}}-150\Delta_2)\times(150\Delta_2,\frac{d_0}{\sqrt{2}}+150\Delta_2,\frac{d_0}{\sqrt{2}}+150\Delta_2)m^3$.
Similar results are observed as in Fig.~\ref{sensing_snr1}. 
It is seen that the sensing error for the single antenna is larger than that for the ULA with the same SNR and $d_0$, because the sensing for a single antenna is in 3-D space while  the sensing for ULA is in 2-D space, and the former is much difficult than the latter.
        }

\subsection{Channel Capacity with ULA Receiver}
\begin{figure}[t]
  \centering
  \centerline{\includegraphics[width=10.7cm,height=8cm]{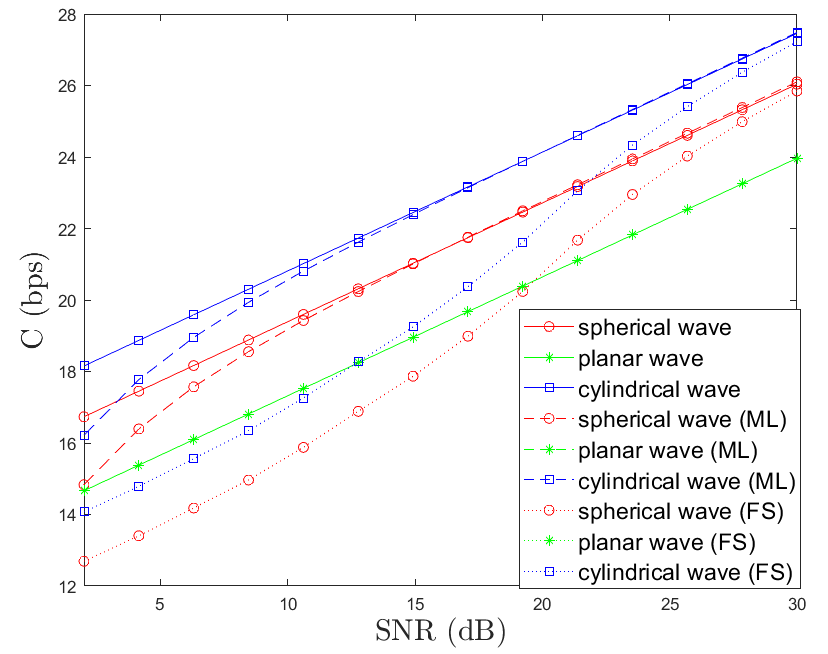}}
  \caption{The channel capacity $C$ versus SNR with ULA.}
  \label{SNRvsC}
\end{figure}

\textcolor{black}{
For the ULA receiver, the power should be concentrated on the line, which cannot be achieved by traditional planar waves. In the design of the reflection coefficient, we target to focus the power at the location of the ULA which is set as $(0m,35m,35m)$.  }

\textcolor{black}{
The channel capacity $C$ versus SNR is shown in Fig.~\ref{SNRvsC}.
The solid lines show the capacities when the location of the receiver is actually known, while the dash lines show the capacities when ML or FS is adopted to sense the location of the receiver. 
It is seen that, $C$ increases with the increase of SNR for all three types of waves, and
the cylindrical waves have the best performance. 
For the capacities without location errors,
the SNR gain of the cylindrical waves over the spherical waves is 4 dB, and the SNR gain of the spherical waves over the planar waves is 6 dB, which demonstrates that the best effect can be achieved only by designing the cylindrical waves for the ULA.  
When SNR increases, the sensing accuracy becomes higher and the degradation of capacities caused by the estimation errors becomes smaller.  Although the estimation errors degrade the capacity for cylindrical and spherical waves, the decrease of capacity for planar waves is negligible. 
 Because the ML sensing method achieves higher accuracy of sensing, the ML method yields higher capacities than the FS method.
}

\begin{figure}[t]
  \centering
  \centerline{\includegraphics[width=10.7cm,height=8cm]{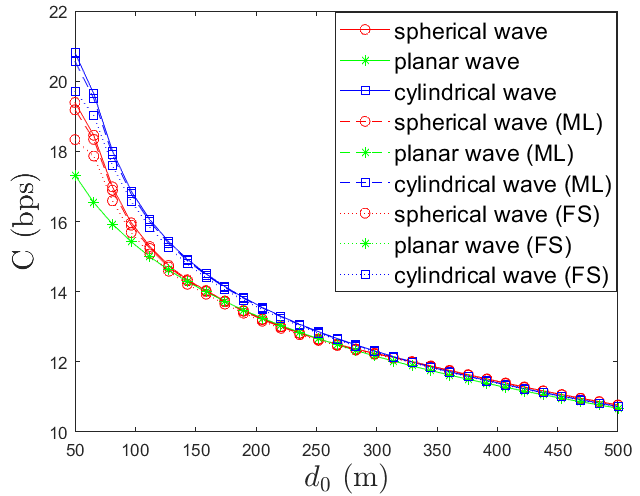}}
  \caption{The channel capacity $C$ versus the distance of the ULA.}
  \label{distance_vs_C_ULA}
\end{figure}

\textcolor{black}{
We further explore the relationship between the channel capacity and the distance of the receiver in Fig.~\ref{distance_vs_C_ULA}, where the transmit power is fixed such that SNR is 10 dB for planar waves at $d_0=50$m. 
 It is seen that, the channel capacities of all three types of waves drop with the increase of the distance. When the distance is above $125$m, the channel capacity of spherical waves is close to that of planar waves. When the distance is above $300$m, the channel capacities of cylindrical, planar and spherical waves  are indistinguishable, which indicates that cylindrical and spherical waves gradually degenerate into planar waves with the increase of the distance. For channel capacities, the superiority of cylindrical waves is only significant within $300$m that is less than half of the Rayleigh Distance. Thus, Rayleigh Distance does not serve as  a good split line for the near and the far field in respect of channel capacity. 
 When the distance increases, the sensing accuracy becomes lower while the influence on the capacities caused by the estimation errors becomes smaller, which explains why capacities do not degrade severely with the increase of the distance.
}

\subsection{Channel Capacity with A Single Antenna  Receiver}
\begin{figure}[t]
  \centering
  \centerline{\includegraphics[width=10.7cm,height=8cm]{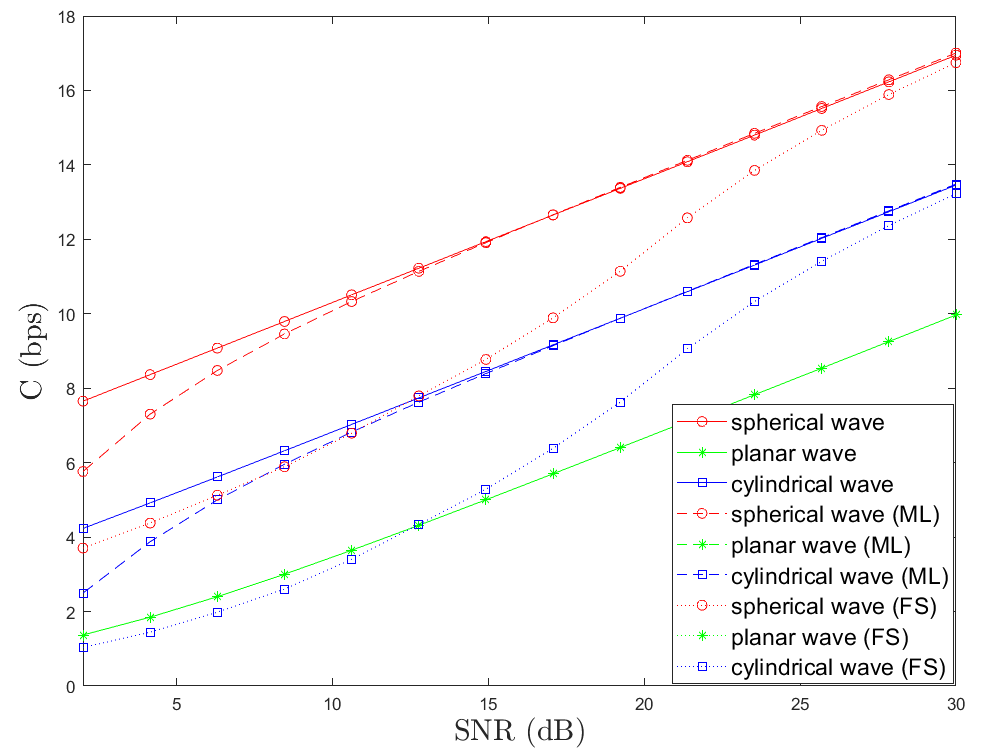}}
  \caption{The channel capacity $C$ versus SNR with the single antenna.}
  \label{SNRvsC_single}
\end{figure}

\textcolor{black}{
When the receiver is equipped with a single antenna, spherical waves would have the best focusing property.  We set the single antenna as the focal point for the spherical waves at $(0m,35m,35m)$.
For the cylindrical waves, since the focusing property is the best when the single antenna is on the focal line, the focal line is set to go through the single antenna. }

\textcolor{black}{
The channel capacity $C$ versus SNR is shown in Fig.~\ref{SNRvsC_single}.
The solid lines show the capacities when the location of the receiver is actually known, while the dash lines show the capacities when ML or FS is adopted to sense the location of the receiver. 
Similar results are observed in Fig.~\ref{SNRvsC}, while the spherical waves have the largest channel capacity.
It is seen that, $C$ increases with the increase of SNR for all three types of waves.
For the spherical waves, $C$ is larger than both the cylindrical waves and the planar waves, because the spherical waves concentrate the majority of the energy on the single antenna. 
The SNR gain of the cylindrical waves over the planar waves is roughly 9 dB, and the SNR gain of the spherical waves over the cylindrical waves is also roughly 9 dB.
It demonstrates that the best effect can be achieved only by designing the spherical waves for the single antenna.
}

\begin{figure}[t]
  \centering
  \centerline{\includegraphics[width=10.7cm,height=8cm]{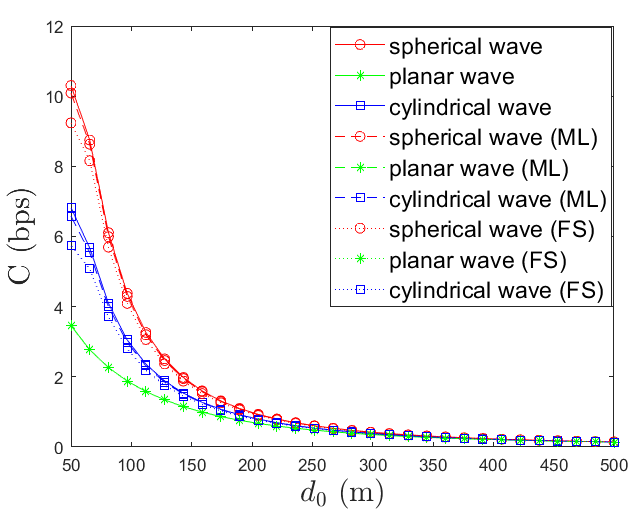}}
  \caption{The channel capacity $C$ versus the distance of the single antenna.}
  \label{distance_vs_C_single}
\end{figure}

\textcolor{black}{
We further explore the relationship between the channel capacity and the distance of the receiver in Fig.~\ref{distance_vs_C_single}, where the transmit power is fixed such that SNR is 10 dB for planar waves at $d_0=50$m. 
The results is similar to Fig.~\ref{distance_vs_C_ULA}, while the spherical waves have the largest channel capacity. When the distance is above $200$m, cylindrical and spherical waves gradually degenerate into planar waves with the increase of the distance. For channel capacities, the superiority of spherical waves is only significant within $200$m that is $1/4$ of the Rayleigh Distance. }
 

\section{Conclusion}
In this paper, we disclose that the conventional planar reflected wave of RIS may cause serious energy leakage when the receiver is located in the near field and is not equipped with planar array. We propose a scheme of RIS's coefficient to convert planar waves into cylindrical waves such that the energy is concentrated on the ULA antenna and a scheme of the RIS to convert planar waves into spherical waves such that the energy is concentrated on the single antenna. Different from the conventional scheme that the planar array receives planar waves, the position of focal line or focal point depends on the position of the receiver and should be obtained before the RIS design. 
\textcolor{black}{
We then propose the ML method and the FS method to sense the location  of the receiver based on the  analytic expression of the reflection coefficient, and derive the corresponding PEB.
Simulation results demonstrate that the proposed scheme can reduce energy leakage and thus enlarge the channel capacity compared to the conventional scheme. Moreover, the location of the receiver could be accurately sensed by the ML method with large computation complexity or be roughly sensed by the FS method with small computation complexity.
}

 \small 
 \bibliographystyle{ieeetr}
 \bibliography{IEEEabrv,RIS}


\ifCLASSOPTIONcaptionsoff
  \newpage
\fi



%

\end{document}